\documentclass[11pt]{article}
\usepackage{latexsym}  
\usepackage{amssymb}
\usepackage{graphicx}
\usepackage{amsmath}

\newcommand{\VEV}[1]{\left<{#1}\right>}

\begin{document}

\begin{center}

{\Large\bf Charged Lepton Mass Relations in a SUSY Scenario }

\vspace{4mm}

\vspace{4mm}
{\bf Yoshio Koide$^a$ and Toshifumi Yamashita$^b$}

${}^{a}$ {\it Department of Physics, Osaka University, 
Toyonaka, Osaka 560-0043, Japan} \\
{\it E-mail address: koide@het.phys.sci.osaka-u.ac.jp}

${}^b$ {\it   Department of Physics, Aichi Medical University, 
Nagakute, Aichi 480-1195, Japan} \\
{\it E-mail address: tyamashi@aichi-med-u.ac.jp }

\date{\today}
\end{center}

\vspace{3mm}
\begin{abstract}
The observed charged lepton masses satisfy the relations 
$K \equiv (m_e +m_\mu+m_\tau)/(\sqrt{m_e} +\sqrt{m_\mu} +\sqrt{m_\tau})^2 =2/3$ and 
$\kappa \equiv \sqrt{m_e m_\mu m_\tau}/(\sqrt{m_e} +\sqrt{m_\mu} 
+\sqrt{m_\tau})^3 =1/486$ 
with great accuracy. 
These parameters are given as $K=( {\rm Tr}[\Phi \Phi])/
 ({\rm Tr}[\Phi ])^2$ and $\kappa = {\rm det} \Phi/
({\rm Tr}[\Phi ])^3$ if the charged lepton masses  
$m_{ei}$ are given by $m_{ei} \propto \sum_k \Phi_i^{\ k} \Phi_k^{\ i}$
where $\Phi$ is a U(3)-family nonet scalar. 
Simple scalar potential forms to realize the relations have been already proposed in 
non-supersymmetric scenarios, but the potential forms are not stable against 
the renormalization group effects. 
In this paper, we examine supersymmetric scenarios 
and find that 
the parameters $K$ and $\kappa$ 
are made
stable against the effects
in a very nontrivial way, 
even though the superpotential itself (in the canonical basis) suffers the usual 
corrections. 
We also show possible simple superpotential forms for the relations.  
\end{abstract}

\vspace{3mm}

\section{Introduction}

It is well known that the charged lepton mass relation \cite{Koidemass}
$$
K \equiv \frac{m_e +m_\mu+m_\tau}{(\sqrt{m_e} +\sqrt{m_\mu} +\sqrt{m_\tau})^2}
 = \frac{2}{3} ,
\eqno(1.1)
$$ 
is satisfied by the observed charged lepton masses (pole masses) as
\cite{PDG16}, 
$$
K(m_{ei}^{obs}) = (2/3)\times(0.999989\pm 0.000014) .
\eqno(1.2)
$$
Naively thinking, this coincidence seems to suggest a nontrivial physics 
behind it. 
In general, however, the ``mass" in the relation derived in a field 
theoretical model means the ``running" mass that is evaluated at a high 
energy scale of the expected nontrivial physics, instead of the ``pole" mass. 
As well known, the renormalization group (RG) evolutions, especially the 
decoupling effects, are family dependent to modify the relation as shown below, 
and thus it is puzzling that the relation holds for the ``pole" masses. 
Since the accuracy is so excellent, it is worth looking for a 
way to make the relation satisfied also by the ``pole" masses, in cases 
that the nontrivial physics exists. 
For this purpose, it is very important to treat the RG effects carefully.

The deviation of $K(m_{ei}^{run})$ from $K(m_{e_i}^{pole})$ is caused by 
the family-dependent logarithmic term $\log(\mu/m_{ei})$ in the radiative 
QED correction to the running mass $m_{ei}(\mu)$ \cite{Arason}
$$
m_{ei}(\mu) = m_{ei}\left\{ 1 - \frac{\alpha(\mu)}{\pi} \left( 
1 + \frac{3}{4} \log \frac{\mu^2}{m_{ei}^{\ 2}} \right) \right\} .
\eqno(1.3)
$$
For this problem, Sumino \cite{Sumino_09} proposed an attractive mechanism 
that introduces family gauge bosons $A_i^{\ j}$ with masses 
$(M^2)_i^{\ i} \propto m_{ei}$:   
the logarithmic term in the radiative QED correction is canceled by
a logarithmic term $\log(\mu/(M^2)_i^{\ i})$ due to the family gauge bosons. 
Then, the mass formula in Eq.~(1.1) is satisfied also by the pole masses, 
assuming that the running masses satisfy the formula above the scale where 
the family gauge bosons are integrated out. 

Recently,  another mass relation for the charged leptons \cite{another_c-mass}
$$
\kappa \equiv \frac{\sqrt{m_e m_\mu m_\tau}}{(
\sqrt{m_e} +\sqrt{m_\mu} +\sqrt{m_\tau})^3}  = \frac{1}{2 \cdot 3^5} = 
\frac{1}{486} ,
\eqno(1.4)
$$
was proposed. 
The relation is satisfied by the observed charged lepton   
masses with accuracy $\sim 10^{-4}$. 

Let us assume that the charged lepton masses 
$m_{ei}$ are given by 
$$
m_{ei} = k_e \sum_k \langle\Phi\rangle_i^{\ k} 
\langle\Phi\rangle_k^{\ i} ,
\eqno(1.5)
$$
where $\langle\Phi\rangle$ is a vacuum expectation value (VEV) 
of a U(3)-family nonet scalar. 
  The form in Eq.~(1.5) is understood from a seesaw scenario 
\cite{K-F_PRD01}.  
Then, the mysterious mass relations in Eqs.~(1.1) and (1.4) can be expressed 
by somewhat intuitive forms 
$$
K= \frac{ {\rm Tr}[\langle\Phi\rangle \langle \Phi \rangle]}{ 
({\rm Tr}[\langle \Phi\rangle ])^2} =\frac23,
\eqno(1.6)
$$
 and 
$$
\kappa = \frac{{\rm det} \langle\Phi\rangle }{
({\rm Tr}[\langle\Phi \rangle])^3} =\frac1{486}. 
\eqno(1.7)
$$ 

 In a non-supersymmetric (non-SUSY) scenario, 
the relation in Eq.~(1.6) has been derived by assuming a scalar potential
 with a simple form  \cite{Koide_MPLA90}
 $$
 V_K = \mu^2 [\Phi \Phi] + \lambda [\Phi \Phi][\Phi \Phi] 
 + \lambda' [\Phi_8 \Phi_8] [\Phi]^2. 
\eqno(1.8)
$$
Here and hereafter, for convenience, we denote Tr$[A]$ as $[A]$ simply.   
The scalar $\Phi_8$ is an octet component of the nonet scalar $\Phi$:
$$
\Phi_8 \equiv  \Phi -\frac{1}{3}  [\Phi] \, {\bf 1} .
\eqno(1.9)
$$
In Ref.~\cite{another_c-mass}, 
the relation in Eq.~(1.7) has been derived by assuming 
another simple scalar potential form
$$
V_\kappa = \lambda' \left( [\Phi_8 \Phi_8 \Phi_8 \Phi_8] 
+  [\Phi_8 \Phi_8 \Phi_8] [\Phi] +  [\Phi_8 \Phi_8 ] [\Phi]^2
+ \frac{1}{3^4} [\Phi]^4 -\frac{1}{4}  [\Phi_8 \Phi_8] [\Phi_8 \Phi_8]
\right) . 
\eqno(1.10)
$$

Once the relations are obtained for the running masses as above,
the Sumino mechanism may ensure the same relations hold also 
for the pole masses. 
However, it is not the end of the story: we should also worry about 
the RG effects at the high energy scale. 
As well known, the scalar potentials in non-SUSY models are not 
stable against the RG evolution. 
This means that, in order to get the above simple forms at the scale 
where the field $\Phi$ is integrated out, the potential forms 
at the cutoff scale must be complicated or fine-tuned.

In this paper, we consider SUSY models to avoid 
the problem and 
give possible superpotential forms which
give $K$ in Eq.~(1.6) and $\kappa$ in Eq.~(1.7). 

We note that, in a SUSY scenario, the original Sumino mechanism does not
work as the vertex corrections with the family gauge boson is suppressed. 
In addition, in the original model, in order to give the minus sign 
for the cancellation, 
the charged leptons $(e_L, e_R)$ are assigned to $({\bf 3}, {\bf 3}^*)$ 
of the family symmetry U(3).
Therefore, the original model is not a conventional U(3) family model 
with no anomaly in the standard model sector. 
In Ref.~\cite{K-Y_PLB12}, we show that 
this problem is avoided  by introducing  
family gauge bosons with inversely hierarchical masses 
$(M^2)_i^{\ i} \propto 1/m_{ei}$, i.e. 
$\log ((M^2)_i^{\ i} )\propto -\log m_{ei}+const$, 
which works in a SUSY scenario, 
although the cancellation holds only approximately in this case.
We assume this modified version of the Sumino mechanism to explain the 
coincidence between $K(m_{ei}^{run})$ and $K(m_{e_i}^{pole})$. 

This paper is organized as follows. 
We give simple superpotentials for the relations of 
$K$ in Eq.~(1.6) in Sec.~\ref{Sec:K} and of $\kappa$ in Eq.~(1.7) 
in Sec.~\ref{Sec:kappa}, respectively. 
In Sec.~\ref{Sec:K}, we discuss the stability of the obtained mass relation 
by applying the discussion given in Ref.~\cite{Borzumati:2009hu} 
for the stability of the effective couplings against the RG effects 
in a context of the SUSY grand unified theory, to our setup. 
The Sec.~\ref{Sec:remarks} is devoted for the concluding remarks.

\vspace{5mm}

\section{$K$ relation in SUSY scenario}
\label{Sec:K}

In this section, we construct a SUSY model for the relation of $K$ in Eq.~(1.6). 

We assume the following superpotential:  
$$
W_K = \frac12\mu_1 \phi_1^{\ 2} +  \frac12\mu_2 \phi_2^{\ 2} + \mu_3 \phi_1 \phi_2 
+ \mu [\Phi\Phi] + \lambda_1 [\Phi_8 \Phi_8]\phi_1 + 
\lambda_2 [\Phi ]^2 \phi_2 ,
\eqno(2.1)
$$
where $\phi_1$ and $\phi_2$ are U(3)-family singlet scalars. 
Then, neglecting the soft SUSY breaking terms, 
we obtain the following three equations:
$$
0 = \frac{\partial W_K}{\partial \phi_1} =  \mu_1 \phi_1+ \mu_3 \phi_2 
+ \lambda_1  [\Phi_8 \Phi_8] ,
\eqno(2.2)
$$
$$
0 = \frac{\partial W_K}{\partial \phi_2} =  \mu_2 \phi_2+ \mu_3 \phi_1 
+ \lambda_2  [\Phi]^2 , 
\eqno(2.3)
$$
$$
0 =  \frac{\partial W_K}{\partial \Phi} = 2 \mu \Phi + 
\lambda_1 \phi_1 \left( 2 \Phi -\frac{2}{3} [\Phi] {\bf 1} \right)
 + \lambda_2 2 \phi_2 [\Phi] {\bf 1} ,
\eqno(2.4)
$$
from the so-called F-flatness condition, while the D-flatness condition makes the VEV 
$\VEV{\Phi_8}$ hermitian. 
Eqs.~(2.2) and (2.3) lead to
$$
\phi_1 = \frac{1}{\mu_3^{\ 2} -\mu_1 \mu_2} 
\left\{\lambda_1 \mu_2 [\Phi_8 \Phi_8] - \lambda_2 \mu_3 
[\Phi ]^2 \right\} ,
\eqno(2.5)
$$
and

$$
\phi_2 = \frac{1}{\mu_3^{\ 2} -\mu_1 \mu_2} 
\left\{\lambda_2 \mu_1 [\Phi]^2 - \lambda_1 \mu_3 
[\Phi_8 \Phi_8] \right\} ,
\eqno(2.6)
$$
respectively.

We assume that $\mu_1$ and $\mu_2$ are negligibly small compared with
$\mu_3$.  
Then, in the limit of ${\mu_1}/{\mu_3} \rightarrow 0$ and 
${\mu_2}/{\mu_3} \rightarrow 0$, we obtain 
$$
\phi_1 = -\frac{\lambda_2}{\mu_3} [\Phi]^2 , \ \ \ \ 
\phi_2 = -\frac{\lambda_1}{\mu_3} [\Phi_8 \Phi_8] . 
\eqno(2.7)
$$

When we substitute the VEVs in Eq.~(2.7) into $F$-flatness condition 
in Eq.~(2.4), we obtain a VEV relation 
$$
 \left( \mu -  \frac{\lambda_1\lambda_2}{\mu_3} [\Phi]^2 \right) \Phi 
-  \frac{\lambda_1\lambda_2}{\mu_3} \left( [\Phi \Phi] 
-\frac{2}{3}  [\Phi]^2 \right) [\Phi] {\bf 1} =0  .
\eqno(2.8)
$$

In order that there is a VEV value $\langle \Phi \rangle \neq {\bf 1} $, 
both the coefficients of $\Phi$ and ${\bf 1}$ must be zero, so that we obtain 
the following relations 
$$
\mu - \frac{\lambda_1\lambda_2}{\mu_3} [\Phi]^2 = 0,
\eqno(2.9)
$$
and
$$
[\Phi \Phi]-\frac{2}{3} [\Phi]^2 =0 . 
\eqno(2.10)
$$
The relation in Eq.~(2.9) plays a role of fixing the scale of VEV of $\Phi$, 
dependently on the parameters $\mu$, $\mu_3$, $\lambda_1$ and $\lambda_2$.  
The VEV relation in Eq.~(2.10) is just the one which we desired.  
Note that the VEV relation is independent of the potential parameters 
$\mu_1$, $\mu_2$, $\mu_3$, $\mu$, $\lambda_1$ and $\lambda_2$.

Here, we give a discussion on the stability of the VEV relations. 
We note that the superpotential in Eq.~(2.1) is not general, as the scalar 
potential in Eq.~(1.8). 
For example, the term $\mu[\Phi\Phi]=\mu([\Phi_8\Phi_8]+\frac13[\Phi]^2)$ is 
a special combination of the two irreducible terms, which have different RG 
evolutions from each other.
Then, at first glance, this SUSY model might seem to suffer from the same 
problem as the non-SUSY models. 
In fact, it is not the case~\cite{Borzumati:2009hu}: the nonrenormalization 
theorem ensures that the 
effective leptonic Yukawa couplings runs in the same way as in the minimal SUSY 
standard model (MSSM).
An important point is that, although the VEV $\VEV\Phi$ actually runs also 
in the SUSY model, the running is caused only by the wave function renormalization 
due to the renormalization of the Kahler potential 
since the superpotential is not renormalized. 
It is straightforward to see that this running of the VEV cancels the wave function 
renormalization of the field $\Phi$ in the coupling $l\Phi\Phi eh_d$, which gives 
effectively the leptonic Yukawa interactions. 
We can understand this cancellation as follows. 
The wave function renormalization is nothing but just the renaming of the fields, which 
does not affect the ``physics" that 
the effective leptonic Yukawa interaction is given by the coupling with the 
``physical" VEV $\VEV\Phi$.
This is analogous to the fact the physical length is independent of the measure. 
In this way, we understand that in the SUSY models, the effective leptonic Yukawa 
coupling runs independently of the running of the VEV $\VEV\Phi$ as far as the 
VEV is determined by the superpotential, which is protected by the nonrenormalization 
theorem. 
This ensures that once we obtain the simple superpotential for the relation of $K$ 
in Eq.~(1.6) at the cutoff scale in some way, 
the relation holds also at the scale where the field $\Phi$ is integrated out, 
even though the superpotential at the scale is corrected to become complicated 
due to the RG evolution. 
Thus, we conclude that in this SUSY model with the help of the modified Sumino mechanism, 
we may explain the relation $K=2/3$ holds for the pole masses, assuming the 
relation is obtained for the running masses at the cutoff scale of the SUSY model.

\vspace{5mm}

\section{The $\kappa$ relation in SUSY scenario}
\label{Sec:kappa}

In this section, we construct a SUSY model for the relation of $\kappa$ in Eq.~(1.7). 

Analogously to the previous section, we assume the following 
superpotential with a simple form
$$
W_\kappa = \mu_{AB} [AB]+ \mu [\Phi\Phi]+ \lambda_A \left\{ [\Phi\Phi A] + 
\alpha [\Phi \Phi] [A] \right\}  + 
 \lambda_B \left\{ [\Phi_8 \Phi_8 B] + 
\beta [\Phi_ 8 \Phi_8] [B] \right\} + \lambda [\Phi\Phi\Phi],
\eqno(3.1)
$$
where $A$ and $B$ are U(3)-family nonet scalars  
and we set the mass terms $\mu_A [AA]$ and $\mu_B [BB]$ 
negligible compared with $\mu_{AB} [AB]$, 
as in the previous section.
The $\lambda_B$ terms can be re-written as 
$$
\lambda_B \left\{ [\Phi\Phi B] -\frac{2}{3} [\Phi] [\Phi B] 
+ \beta [\Phi\Phi][B] + \frac{1}{9}(1 -3 \beta) [\Phi]^2 [B] \right\} .
\eqno(3.2)
$$

We can obtain the following three VEV relations:
$$
0 = \frac{\partial W_\kappa}{\partial A} = \mu_{AB} B + 
\lambda_A ( \Phi\Phi + \alpha [\Phi\Phi ] {\bf 1})  ,
\eqno(3.3)
$$
$$
0 = \frac{\partial W_\kappa}{\partial B} = \mu_{AB} A + 
\lambda_B \left\{ \Phi\Phi 
 -\frac{2}{3} [\Phi] \Phi + \beta [\Phi\Phi ] {\bf 1} 
+\frac{1}{9} (1 -3\beta) [\Phi]^2 {\bf 1} \right\}   ,
\eqno(3.4)
$$
$$
0 = \frac{\partial W_\kappa}{\partial \Phi} = 
2\mu\Phi+3\lambda\Phi\Phi +
\lambda_A \left\{  (\Phi A + A\Phi) 
+2 \alpha [A] \Phi  \right\} 
$$
$$
+  \lambda_B \left\{  (\Phi B + B\Phi) -
\frac{2}{3} [\Phi] B  -\frac{2}{3} [\Phi B] {\bf 1}
+2\beta[B]\Phi
+\frac{2}{9} (1-3\beta) [\Phi][ B] \,  {\bf 1} \right\} . 
\eqno(3.5)
$$
By substituting Eqs.~(3.3) and (3.4) into Eq.~(3.5), 
we obtain a VEV relation
$$
0 = - \frac{\partial W_\kappa}{\partial \Phi} = 
\frac{\lambda_A \lambda_B}{\mu_{AB}} \left\{ 
4   \Phi \Phi \Phi -\frac{4}{3}[\Phi]\Phi \Phi 
+ 4 \left( \alpha+\beta +3 \alpha \beta \right)[\Phi\Phi] \Phi 
\right.
$$
$$
+\left( \frac{2}{9}(1+3 \alpha(1-3 \beta) )-\frac{4}{3} \alpha 
-\frac23\beta
\right) [\Phi]^2 \Phi
- \frac{2}{3}[\Phi \Phi \Phi]\, {\bf 1} 
$$
$$
 \left.
 +\left( - \frac{4}{3}\alpha+ \frac{2}{9}(1+3\alpha)(1-3\beta)  
 \right) [\Phi][\Phi\Phi]\,  {\bf 1}
 \right\} -2\mu\Phi-3\lambda\Phi\Phi. 
\eqno(3.6)
$$

Since, for an arbitrary 3$\times$3 matrix $A$,  we know 
 relations 
$$
AAA = [A] AA + \frac{1}{2} \left( [AA] -[A]^2 \right) A 
+ {\rm det} A \, {\bf 1} ,
\eqno(3.7)
$$ 
and
$$
[AAA] = 3  {\rm det} A + \frac{3}{2}[AA][A] -\frac{1}{2} [A]^3 ,
\eqno(3.8)
$$
we can re-write the relation in Eq.~(3.6) as follows:
$$
0 =\left\{ 2\, {\rm det} \Phi - \frac{1}{27} \left( 5 + 12 (\alpha 
+\beta + 3\alpha \beta) 
\right)  [\Phi]^3 \right\} {\bf 1} + (\Phi \ {\rm and} \ 
\Phi\Phi \ {\rm terms} ). 
\eqno(3.9)
$$
Here, we have already rewritten $[\Phi\Phi]$ as 
$$
[\Phi\Phi]= K_0 + \frac{2}{3} [\Phi]^2 ,
\eqno(3.10)
$$
with $K_0=0$ if the relation in  Eq.~(2.10) is satisfied 
as suggested experimentally.
%

In a similar way as in the previous section, by imposing $\VEV\Phi$ 
is not proportional to the unit matrix, the coefficients of the 
$\Phi\Phi$- and $\Phi$-terms are forced to vanish, which fix the 
scale of the VEV $[\Phi]$ and the VEV relation $K$ as functions of 
the parameters $\lambda$ and $\mu$. 
In this section, we just assume that 
the parameters $\lambda$ and $\mu$ 
are tuned so that an appropriate scale of the VEV $[\Phi]$ and the 
suggested VEV relation $K=2/3$ are 
obtained, 
and concentrate on the coefficient of the unit matrix 
${\bf 1}$ in Eq.~(3.9).  

Thus,  we can finally re-write 
the relation in Eq.~(3.9) only with $[\Phi]^3$ and det$\Phi$ and,
thereby, we obtain a relation on $\kappa$
$$
\kappa =  \frac{{\rm det}\Phi}{[\Phi]^3}  
= +\frac{1}{54} \left\{ 5 +12( \alpha + \beta + 
3 \alpha \beta) \right\} .
\eqno(3.11) 
$$ 
Since we want to reproduce the numerical result in Eq.~(1.7), 
the condition is given by
$$
11 + 27 (\alpha + \beta + 3 \alpha \beta)  = 0.
\eqno(3.12)
$$

We note that, in contrast to the previous section, we have to 
assume a specific relation between the coefficients $\alpha$ 
and $\beta$. 
We suppose that these are given as simple rational numbers 
because we consider that the relation in Eq.~(1.7), 
as well as that in Eq.~(1.6), 
has a connection with the fundamental physics. 
We find simple solutions of the condition in Eq.~(3.12);
$$
(\alpha, \beta) =(-\frac{1}{9},  -\frac{4}{9} ), \ \ \ 
{\rm and} \ \ 
(\alpha, \beta) =(-\frac{4}{9},  -\frac{1}{9} ), 
\eqno(3.13)
$$
i.e. 
$$
W_\kappa  = \mu_{AB} [AB]+ \lambda_A \left\{ [\Phi\Phi A] 
 -\frac{1}{9}  [\Phi \Phi] [A] \right\}  + 
 \lambda_B \left\{ [\Phi_8 \Phi_8 B] -
\frac{4}{9}  [\Phi_ 8 \Phi_8] [B] \right\} ,
\eqno(3.14)
$$
or $W_\kappa$ with $\alpha \leftrightarrow \beta$. 

Although the form of $W_\kappa$ is not general at all, 
the nonrenormalization theorem again protect the effective 
leptonic Yukawa couplings against the RG effect, 
as discussed in the previous section.
\vspace{5mm}

\section{Concluding remarks}
\label{Sec:remarks}

In conclusion, on the basis of SUSY framework, 
we have found superpotential forms 
$W_K$ and $W_\kappa$ with simple coefficients, which 
lead to the charged lepton mass relations   
in Eqs.~(1.1) and (1.4), i.e. those in Eqs.~(1.6) and (1.7), respectively.
Of cause, the potential forms which has been proposed in this 
paper is not unique.  
Those potential forms have been obtained under a guiding principle  
that the coefficients in the potential should be given by 
rational numbers as simple as possible.
At first glance, the relations derived from such tuned superpotentials 
might seem to be destabilized by the RG effects even in the SUSY models.
We showed that, however, the relations are stable against the effects 
by applying the discussion given in Ref.~\cite{Borzumati:2009hu} 
in a context of the SUSY grand unified theory to our set up. 
Thus, we conclude that, once the simple 
superpotential forms are realized (possibly at the cutoff scale) 
in some way in our SUSY models, with the help of the 
modified Sumino mechanism~\cite{K-Y_PLB12}, 
the desired mass relations holds also for the pole masses.

Finally, let us give a brief comment on the phenomenological aspects of 
our models. 
Since our models stand on the modified Sumino mechanism, 
it is predicted that the features discussed in Refs.~\cite{K-Y_PLB12, YK_PLB14} 
will be observed. 
In addition, more detailed analyses on some specific topics, such as 
family gauge bosons with visible lower mass  at the LHC~\cite{K-Y-Y_PLB15} 
and the $\mu$-$e$ conversion~\cite{Koide-Yamanaka_PLB16}, 
are also applied.

 \vspace{5mm}

{\large\bf Acknowledgments} 

The authors are supported by JPS KAKENHI Grant
number JP16K05325.

\vspace{5mm}


\end{document}